\documentclass[aps,preprint,showpacs,showkeys,pra]{revtex4}%
\usepackage{amssymb}
\usepackage{amsfonts}
\usepackage{amsmath}
\usepackage{graphicx}
\usepackage{bm}

\begin{document}
\title{Angular width of Cherenkov radiation with inclusion of multiple scattering: an path-integral approach}
\author{Jian Zheng}
\email{jzheng@ustc.edu.cn}
\affiliation{CAS Key Laboratory of Geospace Environment and Department of Modern Physics, University of Science and Technology of China, Hefei, Anhui 230026, People¡¯s Republic of China}
\affiliation{IFSA Collaborative Innovation Center, Shanghai Jiao Tong University,
Shanghai 200240, People¡¯s Republic of China}

\begin{abstract}
Visible Cherenkov radiation can offers a method of the measurement of the
velocity of a charged particles. The angular width of the radiation is
important since it determines the resolution of the velocity measurement. In
this article, the angular width of Cherenkov radiation with inclusion of
multiple scattering is calculated through the path-integral method, and and the analytical expressions are presented. The condition that multiple scattering process dominates the angular distribution is obtained.

\end{abstract}
\maketitle

\section{Introduction}

The Cherenkov radiation can be emitted when a fast charged particle moves in a
transparent medium with a speed of $v$ greater than the light speed in the
medium. The radiation appears mainly at an angle from the path of the particle
given by $\vartheta_{0}=\arccos[1/(n\beta)]$, where $\beta=v/c$ is the
normalized speed, $c$ is the light speed in vacuum, and $n$ is the refractive
index of the medium. If the speed $v$ is not too close to unity, the measurement
of $\vartheta_{0}$ offers a convenient method for the determination of $v/c$,
and hence the momentum and energy of the particle. Recent experiments shows
that this method can be applied to the study of energetic electrons generated
in ultra-intense laser plasma
interactions~\cite{Brandl2003,Manclossi2006,Habara2010}. Conical forward THz
emission from femosecond-laser-beam filamentation in air is also partially
attributed to Cherenkov radiation by some authors~\cite{Amico2007}. For the
case of simplicity, the motion of electrons is usually assumed
uniform~\cite{Zheng2005}. However, as an electron moves in a medium, it
suffers multiple scattering with the nuclei in medium, and no longer moves
uniformly. Multiple scattering changes the average direction of the particle
and partially destroys the coherence of the radiation along the path, leading
to a wider angular width of the Cherenkov emission cone and a little lower
radiation intensity~\cite{Dedrick1952,Bowler1996,Zheng2013}. Although modern
numerical simulations can be performed to address this topic, an analytical
approach could provide a deeper insight. In this article, based upon the
assumption that the scattering angles between energetic electrons and medium
particles are small, we calculate the effect of multiple scattering on
Cerenkov radiation with the method of path integral~\cite{Laskin1985}. Only
the effects of multiple scattering and diffraction will be considered here.
Analytical formulas of the spectral power of the radiation are derived in
certain circumstances.

\section{Radiation from energetic electrons with inclusion of multiple
scattering}

Radiation from a rapid moving charged particle can be well described with the
classical theory of electrodynamics. The spectral intensity of the radiation
emitted from a moving electron is given by~\cite{Landau1975}
\begin{equation}
\frac{d^{2}\mathcal{E}}{d\omega d\Omega}=\frac{e^{2}\omega^{2}}{(2\pi
)^{2}c^{3}}\int_{0}^{T}dt\int_{0}^{T}dt^{\prime}\quad e^{i\omega(t-t^{\prime
})}\left[  \mathbf{v}(t)\cdot\mathbf{v}(t^{\prime})-\frac{\omega^{2}}{k^{2}%
}\right]  e^{-i\mathbf{k}\cdot\lbrack\mathbf{r}(t)-\mathbf{r}(t^{\prime})]},
\label{radiation intensity}%
\end{equation}
where $\mathbf{v}(t)$ is the velocity, $\mathbf{r}(t)$ is the trajectory, $-e
$ is the electron charge, and $T$ is the time that the electron passes through a
medium which is assumed a transparent slab in our article. The wave frequency
$\omega$ and wave number $k$ in Eq. (\ref{radiation intensity}) satisfy the
dispersion relation of electromagnetic waves in the medium,%
\[
\frac{\omega}{k}=\frac{c}{n},
\]
where $n$ is the refractive index of the transparent slab.

When the scattering angle is small, the velocity of the electron can be
approximately written as,%
\begin{equation}
\mathbf{v}(t)\simeq v\bm{\theta}(t)+v\left[  1-\frac{1}{2}\theta
^{2}(t)\right]  \mathbf{e}_{z}. \label{velocity 1}%
\end{equation}
Here $\bm{\theta}(t)=\theta(x)\mathbf{e}_{x}+\theta_{y}(t)\mathbf{e}_{y}$ is
an two-dimensional random vector describing the scattering angle which is
assumed small, and $\theta^{2}=\theta_{x}^{2}+\theta_{y}^{2}$. In writing Eq.
(\ref{velocity 1}), we assume that the incident electron is along the
$z-$direction, and already neglect the energy loss of the electron in the medium. With aid of Eq.
(\ref{velocity 1}), we have
\begin{equation}
\mathbf{v}(t)\cdot\mathbf{v}(t^{\prime})\simeq v^{2}\left\{
1+\bm{\theta}(t)\cdot\bm{\theta}(t^{\prime})-\frac{\theta^{2}(t)+\theta
^{2}(t^{\prime})}{2}\right\}  , \label{velocity correlation 1}%
\end{equation}
and
\begin{subequations}
\label{phase}%
\begin{align}
&  \mathbf{k}\cdot\left[  \int_{0}^{t}\mathbf{v}(\tau)d\tau-\int%
_{0}^{t^{\prime}}\mathbf{v}(\tau)d\tau\right]  \simeq kvn_{z}(t-t^{\prime
})-\left[  \Phi(t)-\Phi(t^{\prime})\right]  ,\label{phase 1}\\
\Phi(t)  &  =\frac{1}{2}kvn_{z}\int_{0}^{t}\theta^{2}(\tau)d\tau
-kv\mathbf{n}_{\bot}\cdot\int_{0}^{t}\bm{\theta}(\tau)d\tau, \label{phase 2}%
\end{align}
where $\mathbf{n}=\mathbf{n}_{\bot}+n_{z}\mathbf{e}_{z}$ is the unit vector
along the wave vector $\mathbf{k}$, $n_{z}=\cos\vartheta$, and $\vartheta$ is
the angle between the radiation emission and the $z-$axis. Substituting Eqs.
(\ref{velocity correlation 1}) and (\ref{phase}) into Eq.
(\ref{radiation intensity}), the radiation intensity can be written as a
functional of the random scattering angle $\bm{\theta}$,
\end{subequations}
\begin{align}
\frac{d^{2}\mathcal{E}}{d\omega d\Omega}  &  =\frac{e^{2}\omega^{2}v^{2}%
}{(2\pi)^{2}c^{3}}\left(  1-\frac{1}{\beta^{2}n^{2}}\right)  \int_{0}%
^{T}dt\int_{0}^{T}dt^{\prime}\quad e^{i\omega(1-n\beta\cos\vartheta
)(t-t^{\prime})}e^{i\Phi(t)-i\Phi(t^{\prime})}\nonumber\\
&  +\frac{e^{2}\omega^{2}v^{2}}{(2\pi)^{2}c^{3}}\int_{0}^{T}dt\int_{0}%
^{T}dt^{\prime}\quad\left[  \bm{\theta}(t)\cdot\bm{\theta}(t^{\prime})\right]
e^{i\omega(1-n\beta\cos\vartheta)(t-t^{\prime})}e^{i\Phi(t)-i\Phi(t^{\prime}%
)}\nonumber\\
&  -\frac{e^{2}\omega^{2}v^{2}}{2(2\pi)^{2}c^{3}}\int_{0}^{T}dt\int_{0}%
^{T}dt^{\prime}\quad\left[  \theta^{2}(t)+\theta^{2}(t^{\prime})\right]
e^{i\omega(1-n\beta\cos\vartheta)(t-t^{\prime})}e^{i\Phi(t)-i\Phi(t^{\prime}%
)}. \label{radiation intensity 1}%
\end{align}

Equation (\ref{radiation intensity 1}) has to be averaged over the random
variables,
\begin{align}
\frac{d^{2}\mathcal{E}}{d\omega d\Omega}  &  =\frac{e^{2}\omega^{2}v^{2}%
}{(2\pi)^{2}c^{3}}\left(  1-\frac{1}{\beta^{2}n^{2}}\right)  \left\vert
\left\langle \int_{0}^{T}e^{i\omega(1-n\beta\cos\vartheta)t+i\Phi
(t)}dt\right\rangle \right\vert ^{2}\nonumber\\
&  +\frac{e^{2}\omega^{2}v^{2}}{(2\pi)^{2}c^{3}}\left\{  \left\vert
\left\langle \int_{0}^{T}\theta_{x}(t)e^{i\omega(1-n\beta\cos\vartheta
)t+i\Phi(t)}dt\right\rangle \right\vert ^{2}+\left\vert \left\langle \int%
_{0}^{T}\theta_{y}(t)e^{i\omega(1-n\beta\cos\vartheta)t+i\Phi(t)}%
dt\right\rangle \right\vert ^{2}\right\} \nonumber\\
&  -\frac{e^{2}\omega^{2}v^{2}}{(2\pi)^{2}c^{3}}\operatorname{Re}\left\{
\left\langle \int_{0}^{T}\theta^{2}(t)e^{i\omega(1-n\beta\cos\vartheta
)t+i\Phi(t)}dt\right\rangle \left\langle \int_{0}^{T}e^{-i\omega(1-n\beta
\cos\vartheta)t-i\Phi(t)}dt\right\rangle \right\}  ,
\label{radiation intensity 2}%
\end{align}
where $\left\langle \cdots\right\rangle $ denotes the average. In regard to
the random process related to multiple small-angle scattering in an isotropic
medium, the transition probability of the scattering angle $\bm{\theta}$ obeys
the Gaussian distribution,%
\begin{equation}
P(\bm{\theta},t)=\frac{1}{(q\pi t)}e^{-\theta^{2}/qt}.
\label{gaussian distribution 1}%
\end{equation}
The average can then be performed through the path-integral approach
\cite{Laskin1985},%
\begin{equation}
\left\langle f[\bm{\theta}(t)]\right\rangle =\lim_{\substack{N\rightarrow
\infty,\Delta\rightarrow0\\N\Delta=t}}\frac{1}{(q\pi\Delta)^{N/2}}\int
d\bm{\theta}_{1}\cdots d\bm{\theta}_{N}\exp\left\{  -\sum_{n=0}^{N-1}%
\frac{(\bm{\theta}_{n+1}-\bm{\theta}_{n})^{2}}{q\Delta}\right\}
f(\bm{\theta}). \label{path integral}%
\end{equation}
Given the slab thickness of $\ell$, the time $T$ can be considered as a
constant within the range of $qT\ll1,$, i.e.,
\[
T\simeq\frac{\ell}{v}.
\]
We then have
\begin{align}
\frac{d^{2}\mathcal{E}}{d\omega d\Omega}  &  =\frac{e^{2}\omega^{2}v^{2}%
}{(2\pi)^{2}c^{3}}\left(  1-\frac{1}{\beta^{2}n^{2}}\right)  \left\vert
\int_{0}^{T}dte^{i\omega(1-n\beta\cos\vartheta)t}\left\langle e^{i\Phi
(t)}\right\rangle \right\vert ^{2}\nonumber\\
&  +\frac{e^{2}\omega^{2}v^{2}}{(2\pi)^{2}c^{3}}\left\{  \left\vert \int%
_{0}^{T}dte^{i\omega(1-n\beta\cos\vartheta)t}\left\langle \theta
_{x}(t)e^{i\Phi(t)}\right\rangle \right\vert ^{2}+\left\vert \int_{0}%
^{T}dte^{i\omega(1-n\beta\cos\vartheta)t}\left\langle \theta_{y}%
(t)e^{i\Phi(t)}\right\rangle \right\vert ^{2}\right\} \nonumber\\
&  -\frac{e^{2}\omega^{2}v^{2}}{(2\pi)^{2}c^{3}}\operatorname{Re}\left\{
\int_{0}^{T}dte^{i\omega(1-n\beta\cos\vartheta)t+i\Phi(t)}\left\langle
\theta^{2}(t)e^{i\Phi(t)}\right\rangle \int_{0}^{T}dte^{-i\omega(1-n\beta
\cos\vartheta)t}\left\langle e^{i\Phi(t)}\right\rangle dt\right\}  ,
\label{radiation intensity 3}%
\end{align}
It can be shown that [seeing in the Appendix]
\begin{subequations}
\label{average 1}%
\begin{align}
\left\langle e^{i\Phi(t)}\right\rangle  &  =\frac{1}{\cosh\Omega t}%
\exp\left\{  -i\frac{kv_{0}n_{\bot}^{2}}{2n_{z}}\left[  t-\frac{1}{\Omega
}\tanh\Omega t\right]  \right\}  ,\label{average 1a}\\
\left\langle \theta_{x}(t)e^{i\Phi(t)}\right\rangle  &  =\frac{n_{x}}{n_{z}%
}\frac{\cosh\Omega t-1}{\cosh^{2}\Omega t}\exp\left\{  -i\frac{kv_{0}n_{\bot
}^{2}}{2n_{z}}\left[  t-\frac{1}{\Omega}\tanh\Omega t\right]  \right\}
,\label{average 1b}\\
\left\langle \theta_{y}(t)e^{i\Phi(t)}\right\rangle  &  =\frac{n_{y}}{n_{z}%
}\frac{\cosh\Omega t-1}{\cosh^{2}\Omega t}\exp\left\{  -i\frac{kv_{0}n_{\bot
}^{2}}{2n_{z}}\left[  t-\frac{1}{\Omega}\tanh\Omega t\right]  \right\}
,\label{average 1c}\\
\left\langle \theta^{2}(t)e^{i\Phi(t)}\right\rangle  &  =\left[  \frac
{q}{\Omega}\frac{\sinh\Omega t}{\cosh^{2}\Omega t}+\frac{n_{\bot}^{2}}%
{n_{z}^{2}}\frac{(\cosh\Omega t-1)^{2}}{\cosh^{2}\Omega t}\right]
\exp\left\{  -i\frac{kv_{0}n_{\bot}^{2}}{2n_{z}}\left[  t-\frac{1}{\Omega
}\tanh\Omega t\right]  \right\}  , \label{average 1d}%
\end{align}
where
\end{subequations}
\begin{align*}
\Omega &  =\Omega_{0}e^{-i\pi/4},\\
\Omega_{0}  &  =(q\omega)^{1/2}(n\beta/2)^{1/2}\cos^{1/2}\vartheta.
\end{align*}
It is easily to show that the second and third terms on the right hand side of
Eq. (\ref{radiation intensity 3}) is much smaller than the first term in the
case of $q/\omega\ll1$. Therefore, the radiation intensity can be
approximately written as%
\[
\frac{d^{2}\mathcal{E}}{d\omega d\Omega}=\frac{e^{2}v^{2}}{(2\pi)^{2}c^{3}%
}\left(  1-\frac{1}{n^{2}\beta^{2}}\right)  F(\vartheta),
\]
where the function $F(\vartheta)$ describes the angular distribution of the
radiation%
\begin{equation}
F(\vartheta)=\left\vert \int_{0}^{T\omega}d\xi\frac{\exp\left\{
i(1-n\beta\cos\vartheta)\xi-i\frac{(n\beta/2)\sin^{2}\vartheta}{\cos\vartheta
}\left[  \xi-\frac{\omega}{\Omega_{0}}e^{i\pi/4}\tanh\left[  e^{-i\pi
/4}(\Omega_{0}/\omega)\xi\right]  \right]  \right\}  }{\cosh\left[
e^{-i\pi/4}(\Omega_{0}/\omega)\xi\right]  }\right\vert ^{2}.
\label{angular distribution}%
\end{equation}

In real experiment, we usually have
\[
T\omega=\frac{\ell}{v}\frac{2\pi c}{\lambda}=\frac{2\pi}{\beta}\frac{\ell
}{\lambda}\gg1,
\]
where $\lambda$ is the wavelength of the radiation in vacuum, and $\beta=v/c$ is the
normalized charge velocity. Usually, we also have the condition%
\[
\frac{q}{\omega}\ll1.
\]
The ratio of $\Omega_{0}$ to $\omega$ is then a small number,%
\[
(\Omega_{0}/\omega)=(q/\omega)^{1/2}(n\beta/2)^{1/2}\cos^{1/2}\vartheta\ll1.
\]
In this case, the amplitude of the integrand in Eq.
(\ref{angular distribution}) decays to the half of its maximum around
\begin{equation}
\xi_{0}=\frac{3^{1/3}}{(q/\omega)^{1/3}(n\beta)^{1/6}\cos^{1/6}\vartheta}.
\label{critical point}%
\end{equation}
If the parameter $\xi_{0}$ satisfies the condition
\begin{equation}
\xi_{0}\gg T\omega, \label{condition 1}%
\end{equation}
we can make the appromimation%
\begin{align*}
F(\vartheta)  &  \simeq\left\vert \int_{0}^{T\omega}d\xi\exp\left\{
i(1-n\beta\cos\vartheta)\xi\right\}  \right\vert ^{2}\\
&  =4\frac{\sin^{2}[(1-n\beta\cos\vartheta)T\omega/2]}{(1-n\beta\cos
\vartheta)^{2}}.
\end{align*}
We return to the collisionless limit, in which the angular width is%
\begin{equation}
(\Delta\vartheta)_{\text{diff}}\sim\frac{1}{(T\omega)n\beta\sin\vartheta_{0}}.
\label{angular width 2}%
\end{equation}

When the parameter $\xi_{0}$ is much smaller than the upper integral limit of
Eq. (\ref{angular distribution}), i.e.,
\begin{equation}
\xi_{0}\ll T\omega,\label{condition 2}%
\end{equation}
or
\[
\frac{3^{1/3}}{(q/\omega)^{1/3}(n\beta)^{1/6}\cos^{1/6}\vartheta}\ll T\omega
\]
the function $F(\vartheta)$ can be approximated as
\begin{align}
F(\vartheta) &  \simeq\left\vert \int_{0}^{\infty}d\xi\exp\left\{
i(1-n\beta\cos\vartheta)\xi-\frac{(q/\omega)(n\beta/2)^{2}\sin^{2}\vartheta
}{3}\xi^{3}\right\}  \right\vert ^{2}\nonumber\\
&  =\frac{1}{[(q/\omega)(n\beta/2)^{2}\sin^{2}\vartheta]^{2/3}}\left\vert
\int_{0}^{\infty}d\xi\exp\left\{  i\frac{(1-n\beta\cos\vartheta)}%
{[(q/\omega)(n\beta/2)^{2}\sin^{2}\vartheta]^{1/3}}\xi-\frac{1}{3}\xi
^{3}\right\}  \right\vert ^{2}.\label{angular distribution 1}%
\end{align}
In this case, the function $F(\vartheta)$ is independent of the slab
thickness, indicating that multiple scattering processes dominate the angular
distribution of the radiation, and that only part of the electron trajectory
can contribute the observed radiation. Introducing the function
\begin{equation}
H(x)=\left\vert \int_{0}^{\infty}e^{ixt-t^{3}/3}dt\right\vert ^{2}%
,\label{definition H}%
\end{equation}
the angular distribution function can be written as%
\begin{equation}
F(\vartheta)=\frac{1}{[(q/\omega)(n\beta/2)^{2}\sin^{2}\vartheta]^{2/3}%
}H\left\{  \frac{(1-n\beta\cos\vartheta)}{[(q/\omega)(n\beta/2)^{2}\sin
^{2}\vartheta]^{1/3}}\right\}  .\label{angular distribution 2}%
\end{equation}
\begin{figure}[ptb]
\centering
\includegraphics[width=60mm]{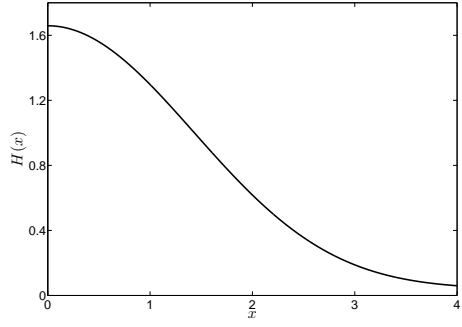}\caption{The function
$H(x)$ versus the augment $x$.}%
\label{function H}%
\end{figure}
We plot the function $H(x)$ in Fig.~\ref{function H}. As seen in this
figure, the maximum of the function $H(x)$ locates at $x=0$, indicating that
the radiation emission reaches its peak at the angle
\[
\vartheta_{0}=\arccos(1/n\beta).
\]
This result means that multiple scattering nearly does not shift the
peak position of the Cherenkov emission. The full width at the half maximum of
the peak is about,
\begin{equation}
(\Delta\vartheta)_{\text{ms}}\simeq\frac{(q/\omega)^{1/3}}{\sin^{1/3}%
\vartheta_{0}(n\beta)^{1/3}}.\label{angular width 1}%
\end{equation}
Under the condition of Eq. (\ref{condition 2}), this angular width is
significantly greater than that in the collisionless limit.

The spectral power of the Cherenkov radiation with inclusion of multiple
scattering is already calculated through the path integral method
\cite{Zheng2013}. In the case of
\begin{equation}
\frac{n\beta-1}{\sqrt{q/\omega}}\gg1, \label{condition 3}%
\end{equation}
The spectral power is approximately given by,
\begin{equation}
\frac{dP}{d\omega}=\frac{e^{2}\omega\beta}{cn}\left[  1-(n\beta)^{-2}\right]
-\frac{2e^{2}q}{3\pi cn^{3/2}[1-(n\beta)^{-2}]}. \label{spectral power 1}%
\end{equation}
We can see that the reduction of the Cherenkov radiation due to multiple
scattering is proportional to the ratio of $q/\omega$, which is usually very
small. In the limit of
\[
\left\vert \frac{\beta n-1}{\sqrt{q/\omega}}\right\vert \ll1,
\]
the so-called Landau-Pomerachuk-Migdal effect plays important role
\cite{Landau1953,Migdal1956},%
\[
\frac{dP}{d\omega}=\frac{e^{2}(q\omega)^{1/2}}{\pi cn^{5/2}}.
\]
In this case, the radiation is essentially due to bremsstrahlung
\cite{Klein1999}.

\section{Discussion and Summary}

We calculate the effect of multiple scattering on the Cherenkov radiation
through the path-integral method. For the sake of simplicity, we make some
important assumptions in our calculation. The first is the neglect of the
energy loss of the charged particles, the second is the Gaussian transition
probability of the scattering angle, seeing Eq. (\ref{gaussian distribution 1}%
), and the third is that the root of mean square of the scattering angle is
much less than one. These assumptions lead to a slight different results of
ours from previous studies like Dedrick's~\cite{Dedrick1952} and
Bowler's~\cite{Bowler1996}. However, the main conclusions are the same.

\begin{acknowledgments}
This work is supported by the Natural Science Foundation of China (Grant Nos.
11175179 and 11475171).
\end{acknowledgments}

\appendix*

\section{The calculation of the path integral}

We briefly present the calculation of the path integral. Denoting%
\begin{equation}
\phi=i\alpha\int_{0}^{t}\theta^{2}(\tau
)d\tau+i\beta\int_{0}^{t}\theta(\tau)d\tau,\label{phase A}%
\end{equation}
where $\theta(t)$ is an one-dimensional random variable whose transition
probability is a gaussian, we have
\begin{align}
I &  \equiv\left\langle e^{\mu\theta^{2}(t)+\eta\theta(t)+i\phi}\right\rangle
=\lim_{\substack{\Delta\rightarrow0,N\rightarrow\infty\\\Delta N=t}}\frac
{1}{(\pi q\Delta)^{N/2}}\int d\theta_{1}\cdots\int d\theta_{N}\nonumber\\
&  \exp\left[  -\sum_{n=0}^{N-1}\frac{(\theta_{n+1}-\theta_{n})^{2}}{q\Delta
}+\mu\theta_{N}^{2}+\eta\theta_{N}+i\alpha\Delta\sum_{n=1}^{N}\theta_{n}%
^{2}+i\beta\Delta\sum_{n=1}^{N}\theta_{n}\right]  ,\label{path integral A1}%
\end{align}
where $\Delta=t/N$. It is easy to see that%
\[
\left\langle e^{i\phi}\right\rangle =I_{\mu=0,\eta=0},
\]
\[
\left\langle \theta(t)e^{i\phi}\right\rangle =\left(  \frac{\partial
I}{\partial\eta}\right)  _{\mu=0,\eta=0},
\]%
\[
\left\langle \theta^{2}(t)e^{i\phi}\right\rangle =\left(  \frac{\partial
I}{\partial\mu}\right)  _{\mu=0,\eta=0}.
\]

After some manipulations, we have%
\[
\left\langle e^{i\phi}\right\rangle =\lim_{\substack{\Delta\rightarrow
0,N\rightarrow\infty\\\Delta N=t}}\frac{1}{\pi^{N/2}}\int d\theta_{1}%
\cdots\int d\theta_{N}\exp\left\{  -\sum_{l,m=1}^{N}\theta_{l}A_{lm}\theta
_{m}+2\sum_{m=1}^{N}B_{m}\theta_{m}\right\}
\]
where the matrix elements $A_{nm}$ is given by
\begin{align*}
A_{N,N} &  =1-\mu q\Delta-i\alpha q\Delta^{2},\\
A_{n,n} &  =2-i\alpha q\Delta^{2},\text{ }(1\leqslant n\leqslant N-1),\\
A_{n,n+1} &  =A_{n,n-1}=-1,\\
A_{n,m} &  =0,\text{ otherwise},
\end{align*}
and the vector component $B_{m}$ is%
\begin{align*}
B_{N} &  =\frac{1}{2}\eta q^{1/2}\Delta^{1/2}+\frac{i}{2}\beta q^{1/2}%
\Delta^{3/2},\\
B_{n} &  =\frac{i}{2}\beta q^{1/2}\Delta^{3/2},\text{ }(1\leqslant
n\leqslant N-1).
\end{align*}
The integral can be accomplished with variable transformation,%
\begin{equation}
I =\lim_{N\rightarrow\infty}\frac{1}%
{\sqrt{\det A}}\exp\left(  \sum_{l,m=0}^{N}B_{l}A_{lm}^{-1}B_{m}\right)
,\label{path integral A2}%
\end{equation}
where $A_{lm}^{-1}$ is the element of the inverse matrix of $A$. The
calculations of $\sqrt{\det A}$ and $\sum_{l,m=0}^{N}B_{l}A_{lm}^{-1}B_{m}$ can be
found in the article by E. W. Montroll~\cite{Montroll1952}. The results
are%
\begin{equation}
\lim_{N\rightarrow\infty}\det A=D(\tau=0)=\cosh(\Omega t)-\frac{\mu q}{\Omega
}\sinh(\Omega t),\label{determinant 1}%
\end{equation}%
and
\begin{align}
\lim_{N\rightarrow\infty}\sum_{l,m=0}^{N}B_{l}A_{lm}^{-1}B_{m}= &
-\frac{\beta^{2}q}{4}\int_{0}^{t}\frac{d\tau}{D^{2}(\tau)}\left(  \int_{\tau
}^{t}D(\tau^{\prime})d\tau^{\prime}\right)  ^{2}\nonumber\\
&  +\frac{i}{2}\eta\beta q\int_{0}^{t}d\tau D(\tau)\int_{0}^{\tau}\frac
{d\tau^{\prime}}{D^{2}(\tau^{\prime})}\nonumber\\
&  +\frac{1}{4}\eta^{2}q\int_{0}^{t}\frac{d\tau}{D^{2}(\tau)},\label{phase B}%
\end{align}
where the function $D(\tau)$ is defined as%
\begin{equation}
D(\tau)=\cosh[\Omega(t-\tau)]-\mu\frac{q}{\Omega}\sinh[\Omega(t-\tau
)]\label{function D}%
\end{equation}
and
\[
\Omega^{2}=-i\alpha q.
\]
With these results, we have%
\begin{subequations}
\label{average 2}%
\begin{align}
\left\langle e^{i\phi}\right\rangle  & =\frac{1}{\sqrt{\cosh\Omega t}}%
\exp\left\{  -\frac{\beta^{2}q}{4\Omega^{3}}\left(  \Omega t-\tanh\Omega
t\right)  \right\}  ,\label{average 2a}\\
\left\langle \theta(t)e^{i\phi}\right\rangle  & =\frac{i}{2}\frac{\beta
q}{\Omega^{2}}\left(  1-\frac{1}{\cosh\Omega t}\right)  \frac{1}{\cosh
^{1/2}\Omega t}\exp\left\{  -\frac{\beta^{2}q}{4\Omega^{3}}\left(  \Omega
t-\tanh\Omega t\right)  \right\}  \label{average 2b}\\
\left\langle \theta^{2}(t)e^{i\phi}\right\rangle  & =\left[  \frac{q}{2\Omega
}\frac{\sinh\Omega t}{\cosh^{3/2}\Omega t}-\frac{\beta^{2}q^{2}}{4\Omega^{4}%
}\frac{(\cosh\Omega t-1)^{2}}{\cosh^{5/2}\Omega t}\right]  \exp\left\{
-\frac{\beta^{2}q}{4\Omega^{3}}\left(  \Omega t-\tanh\Omega t\right)
\right\}  .\label{average 2c}%
\end{align}
\end{subequations}

\end{document}